\documentclass[pra,twocolumn,showpacs]{revtex4}
\usepackage{amsmath}
\usepackage{amssymb}
\usepackage{epsfig}

\begin{document}

\title{\bf Dynamics of quantum vortices in a quasi-two-dimensional Bose-Einstein condensate with two ``holes''}
\author{V.~P. Ruban}
\email{ruban@itp.ac.ru}
\affiliation{L.D. Landau Institute for Theoretical Physics RAS, Moscow, Russia} 

\date{\today}

\begin{abstract}
The dynamics of interacting quantum vortices in a quasi-two-dimensional spatially inhomogeneous Bose-Einstein
condensate, whose equilibrium density vanishes at two points of the plane with a possible presence
of an immobile vortex with a few circulation quanta at each point, has been considered in a hydrodynamic
approximation. A special class of density profiles has been chosen, so that it proves possible to calculate
analytically the velocity field produced by point vortices. The equations of motion have been given in a 
noncanonical Hamiltonian form. The theory has been generalized to the case where the condensate forms 
a curved quasi-two-dimensional shell in the three-dimensional space.
\end{abstract}

\maketitle

\section{Introduction}

Quantum vortices and their dynamics as an intrinsic part of physics of Bose-Einstein condensed gases
have been studied in numerous works (see [1-5] and references therein). Quasi-two-dimensional Bose 
systems at zero temperature deserve special attention, as the theory of vortices in these systems can 
be relatively simple. The wavefunction of a condensate pressed by an external three-dimensional potential 
to the ${\bf x}=(x,y)$  plane after splitting out the transverse coordinate, the motion along which is frozen, 
obeys the two-dimensional Gross-Pitaevskii equation (possibly, with an inhomogeneous nonlinear coefficient).
Here, vortices, which are zeroes of the wavefunction, are just points on the plane, in contrast to vortex lines
in three-dimensional condensates. The velocity circulation around each vortex is $\Gamma=2\pi\hbar/m_{\rm atom}$.
Mathematically similar objects have been long known also in nonlinear optics ([6] and references therein). Vortex
dynamics usually occurs against a spatially inhomogeneous background of the equilibrium two-dimensional density 
$\rho({\bf x})$, whose profile in the case of a Bose gas is determined by the trap potential. A sufficiently
large quasi-two-dimensional condensate in the Thomas-Fermi regime can be described to a good
accuracy by the classical hydrodynamic theory of shallow water above an uneven bottom with the depth
$h({\bf x})\propto\rho({\bf x})$. To build a compact theory of interacting vortices, it is important that, 
in the limit of large distances between the vortices and low velocities of their motion, 
when the potential excitations are negligible, the motion of ``point'' vortices in the classical 
hydrodynamics obeys noncanonical Hamiltonian equations of the form [7, 8]
\begin{equation}
 \Gamma \sigma_n \rho({\bf x}_n)\dot x_n=\frac{\partial H}{\partial y_n},\qquad
-\Gamma \sigma_n \rho({\bf x}_n)\dot y_n=\frac{\partial H}{\partial x_n},
\label{dot_xy}
\end{equation}
where $\sigma_n=+1$ and $-1$ for the positive and negative orientations of the $n$-th vortex, respectively, and 
$H(\{{\bf x}_n\})$ is the Hamiltonian (kinetic energy) of the system depending on the positions of all vortices. 
In addition, the velocity field satisfies the condition $\mbox{div}(\rho{\bf V})=0$,
so that in the two-dimensional space $\rho V_x=\partial_y\psi$ and $\rho V_y=-\partial_x\psi$, where
$\psi({\bf x})$ is the flux potential (stream function). 
The vorticity of the flow is  $\omega=\partial_x v_y- \partial_y v_x$;
i.e., the functions $\psi$ and $\omega$  are related by the linear equation
\begin{equation}
-\nabla_{\bf x}\cdot\frac{1}{\rho({\bf x})}\nabla_{\bf x} \psi({\bf x})=\omega({\bf x}).
\end{equation}
The vorticity of the system of ``point'' vortices is
$\omega({\bf x})\approx \Gamma\sum_n\sigma_n\delta({\bf x}-{\bf x}_n)$.
Thus, the necessity of calculating the Green's function naturally arises, which
means solving the equation [1-4, 6, 8]
\begin{equation}\label{Green_eq_general}
-\nabla_{\bf x}\cdot\frac{1}{\rho({\bf x})}\nabla_{\bf x} G({\bf x},{\bf x}_0)=2\pi\delta({\bf x}-{\bf x}_0).
\end{equation}
The Hamiltonian of vortices can be expressed in a simple form in terms of $G({\bf x},{\bf x}_0)$. 
Actually, if the density does not vanish anywhere,
\begin{eqnarray}
H&=&\frac{1}{2}\int \rho{\bf V}^2  d^2{\bf x}
=\frac{1}{2}\int \frac{(\nabla_{\bf x}\psi)^2}{\rho({\bf x})} d^2{\bf x}
=\frac{1}{2}\int \psi\omega  d^2{\bf x}\nonumber\\
&=&\frac{\Gamma^2}{4\pi}\sum_n F({\bf x}_n)
+\frac{\Gamma^2}{4\pi}{\sum_{m, n}}'\sigma_n\sigma_m G({\bf x}_n,{\bf x}_m).
\end{eqnarray}
Here, the prime sign at the double sum implies that the terms with $m=n$ must be omitted. 
The diagonal terms should be regularized owing to the logarithmic
divergence of the two-dimensional Green's function
at close arguments: $F({\bf x}_n)= G({\bf x}_n,{\bf x}_n+{\bf e}\xi({\bf x}_n))$, where $\xi({\bf x})$ 
is the width of the vortex core, presumably small compared to the typical inhomogeneity distance $R_*$.
The size $\xi$ in the theory of shallow water and in the Gross-Pitaevskii equation (if the effective nonlinear
coefficient is constant) depends in the same way on the local density/depth:
$\xi({\bf x})=\xi_*[\rho_0/\rho({\bf x})]^{1/2}, \quad \Lambda=\log( R_*/\xi_*)\gg 1$.

Thus, the problem of calculating the Hamiltonian of the system of point vortices is reduced to solving Eq.(3).
Substituting there $G({\bf x},{\bf x}_0)=\sqrt{\rho({\bf x})}\sqrt{\rho({\bf x}_0)} g({\bf x},{\bf x}_0)$, we
come to the equation of the form [1, 3, 6]
\begin{equation}
[-\nabla^2_{\bf x}+\tilde\kappa^2({\bf x})] g({\bf x},{\bf x}_0)=2\pi \delta({\bf x}-{\bf x}_0),
\end{equation}
where $\tilde\kappa^2({\bf x})=\sqrt{\rho}\nabla^2_{\bf x}({1}/{\sqrt{\rho}})$. In the majority of cases
with a few exceptions, the solutions cannot be expressed in an explicit form. In particular, if
$\tilde\kappa^2({\bf x})=\mbox{const}=\kappa^2$, the possible density profiles are given by the expression
\begin{equation}
\rho_\kappa=\Big[\int_0^{2\pi}C(\varphi)\exp(\kappa x\cos\varphi+\kappa y\sin\varphi)\frac{d\varphi}{2\pi}\Big]^{-2},
\label{rho_kappa}
\end{equation}
where $C(\varphi)$  is an arbitrary nonnegative function. Each function $\rho_\kappa$ has a single maximum 
and does not vanish at finite points. The Green's function is
\begin{equation}
G_\kappa({\bf x}_1,{\bf x}_2)=\sqrt{\rho_\kappa({\bf x}_1) \rho_\kappa({\bf x}_2)}
K_0(\kappa|{\bf x}_1-{\bf x}_2|),
\label{G_kappa}
\end{equation}
where $K_0(\dots)$ is the corresponding modified Bessel function. In addition to this class of density profiles,
the exact Green's function was found for the linear dependence ($\rho=x$ at $x>0$) [3] and for the Gaussian
density profile [8].

This work is aimed at considering one more, previously unexplored family of the dependences $\rho({\bf x})$ for
which the Green's function can be found exactly. In particular, in a certain region of parameters, it is a
condensate forming an inhomogeneous planar disk with a hole (i.e., a ring). Since the density in this case
vanishes not only at infinity, there is an opportunity of placing a few circulation quanta at each finite zero
point and studying in detail their influence on the vortex motion. In other words, the approach developed in
this work is applicable to a wide class of topical problems associated with so-called giant vortices ([9-14]
and references therein).

In addition, the theory will be generalized to the case of a quasi-two-dimensional Bose-Einstein 
condensate which is not flat but is situated on a curved surface in the three-dimensional space (e.g., on a
sphere, ellipsoid, or toroid [15-18]). An analog of this situation in classical hydrodynamics is the motion of
vortices in an ocean of variable depth on a spherical planet. In contrast to thoroughly studied dynamics of
vortices in a homogeneous liquid layer on curved surfaces ([19-24] and numerous references therein), a
self-consistent theory of vortices in an inhomogeneous curved layer has not yet been developed.

\section{Model of a ``double-hole'' condensate}

At this point, we turn to a qualitatively more complicated situation than that described by Eq.(6), when
the condensate density vanishes at two points of the extended plane (the plane itself plus an infinite point).
To have an example convenient for analytical and numerical study, we introduce the complex variable
$z=x+iy$ and consider the following special case of density profiles with two ``holes'':
\begin{equation}
\rho({\bf x})= 4\Big(\Big|\frac{Az}{1-Bz}\Big|^{\alpha}+ \Big|\frac{1-Bz}{Az}\Big|^{\alpha}\Big)^{-2}, 
\end{equation}
with the parameters $\alpha > 0$,  $A > 0$, and $B\ge 0$. We introduce the curvilinear conformal coordinates
${\bf u}=(u,v)$ in the plane ${\bf x}$ by the analytic function
\begin{equation}
u+iv\equiv w=\ln\left({Az}/[1-Bz]\right).
\end{equation}
The inverse mapping is specified by the expression
\begin{equation}
z=1/[Ae^{-w}+B],
\end{equation}
and the coordinates $(u,v)$  are situated on a cylinder: $u\in(-\infty:+\infty),\, v\in[0:2\pi)$. 
The Jacobian of this mapping is 
\begin{equation}
J({\bf u})=|{Ae^{-w}}/{(Ae^{-w}+B)^2}|^2.
\end{equation}
The density in terms of the new coordinates is given by the formula
\begin{equation}
\rho({\bf u})= \rho(u)={1}/{\cosh^2(\alpha u)}.
\end{equation}
The dimensionless equations of motion of vortices in terms of ${\bf u}_n$ have the form
\begin{eqnarray}
 \sigma_n J({\bf u}_n)\rho(u_n)\dot u_n&=&\partial \tilde H_\alpha/\partial v_n,
\label{eq_u}\\
-\sigma_n J({\bf u}_n)\rho(u_n)\dot v_n&=&\partial \tilde H_\alpha/\partial u_n.
\label{eq_v}
\end{eqnarray}

The equation for the Green's function in the conformal variables preserves its simple structure:
\begin{equation}
-\nabla_{\bf u}\cdot\frac{1}{\rho( u)}\nabla_{\bf u} G({\bf u},{\bf u}_0)=2\pi\delta({\bf u}-{\bf u}_0).
\end{equation}
It is supplemented by the $2\pi$-periodic boundary conditions with respect to the coordinate $v$, a zero 
asymptotics at  $u\to+\infty$,  and the condition of a zero velocity circulation around the point $z=0$, which implies
\begin{equation}\label{condition_u}
\lim_{u\to-\infty}\frac{\partial_u G({\bf u},{\bf u}_0)}{\rho(u)} =0.
\end{equation}
The solution is found by substituting
\begin{equation}
G({\bf u},{\bf u}_0)={g_\alpha({\bf u},{\bf u}_0)}/[{\cosh(\alpha u)\cosh(\alpha u_0)}].
\end{equation}
The function $g_\alpha({\bf u},{\bf u}_0)$ satisfies the equation with constant coefficients,
\begin{equation}
\label{g_alpha}
[-\nabla^2_{\bf u} +\alpha^2] g_\alpha({\bf u},{\bf u}_0)=2\pi \delta({\bf u}-{\bf u}_0).
\end{equation}
The sought solution is
\begin{equation}
 g_\alpha({\bf u},{\bf u}_0)=\mathsf{g}(u-u_0,v-v_0)+\frac{1}{2\alpha}e^{-\alpha(u+u_0)},
\end{equation}
where the function $\mathsf{g}(U,V)$ even in both of its variables is given by the rapidly converging sum
\begin{equation}
\mathsf{g}(U,V)=\!\sum_{l=-\infty}^{+\infty}\!K_0\Big(\alpha\sqrt{U^2+(V+2\pi l)^2}\Big).
\end{equation}

It is worth noting that, at large $|U|$  values, the approximate equality 
$\mathsf{g}(U,V)\approx\exp(-\alpha|U|)/2\alpha$ holds, which allows easily verifying the fulfillment 
of condition (16). On the contrary, at small arguments, we find
$\mathsf{g}(U,V)\approx \ln(1/\sqrt{U^2+V^2}) + \mbox{const}$.

\section{Hamiltonian}

Composing the Hamiltonian of vortices with the use of the found Green's function, one can take into
account the presence of an arbitrary integer number $Q_0$ of circulation quanta around the point $z=0$ by the
respective modification of the asymptotic condition for the flux function at $u\to-\infty$. It should also be
mentioned that at $B\neq 0$  several ($|Q_\infty|$) vortices can be placed near the point 
$w_\infty=\ln(A/B) +i\pi$ , which corresponds to $z=\infty$. Since the Jacobian $J$ diverges at
this point, equations of motion (13) and (14) imply that these vortices will remain immobile at the point
$(u_\infty,\pi)$. This way, arbitrary (integer) numbers $Q_0$ and $Q_1$ of circulation quanta around each of 
two finite zero-density points $z_0=0$ and $z_1=1/B$ can be provided, because the integers 
$Q_0$, $Q_1$, and $Q_\infty$ satisfy the relation
\begin{equation}
Q_0 + Q_1+ Q_\infty +\sum_{n=1}^N\sigma_n=0,
\end{equation}
where $N$ is the number of mobile vortices. As a result, the dimensionless Hamiltonian is given by the expression
\begin{eqnarray}
&&\tilde H_\alpha^{\{N\}}=\frac{1}{2}\sum_{n}
\frac{\Lambda_0+\ln[\sqrt {(J({\bf u}_n)/J_0)}/\cosh(\alpha u_n)]}{\cosh^2(\alpha u_n)} \nonumber\\
&&\qquad+\frac{1}{\alpha}\Big\{Q_0+\frac{Q_\infty}{(1+e^{2\alpha u_\infty})}
+\sum_{n}\frac{\sigma_n}{(1+e^{2\alpha u_n})}\Big\}^2\nonumber\\
&&\qquad+{\sum_{n,m}}'\frac{\sigma_n\sigma_m}{2}
\frac{\mathsf{g}(u_n\!-\!u_m,v_n\!-\!v_m)}{\cosh(\alpha u_n)\cosh(\alpha u_m)}\nonumber\\
&&\qquad+Q_\infty\sum_{n}\sigma_n\frac{\mathsf{g}(u_n\!-\!u_\infty,v_n\!-\!\pi)}
{\cosh(\alpha u_n)\cosh(\alpha u_\infty)},
\label{H_alpha}
\end{eqnarray}
where $\Lambda_0=\mathsf{g}(\xi_*/\sqrt {J(0)},0)\gg 1$ is a large logarithm. The applicability of this formula 
requires that all numerators in the first sum should not be small compared to unity. In the opposite case, the
approximation of a point vortex fails.

In all numerical examples presented below, we set $\Lambda_0=7.0$ and $\alpha=2.0$, and all vortices have a 
positive sign, i.e., $\sigma_n=+1$. 

To see a qualitative difference of this system from systems of type (6), we first consider the dynamics of a
single vortex. The trajectories of the vortex in the plane are the level contours of the Hamiltonian (22) at
$N=1$. Two examples of phase portraits with different sets of parameters $A$, $B$, $Q_0$, and $Q_1$ are shown 
in Fig.1.

\begin{figure}
\begin{center}
 \epsfig{file=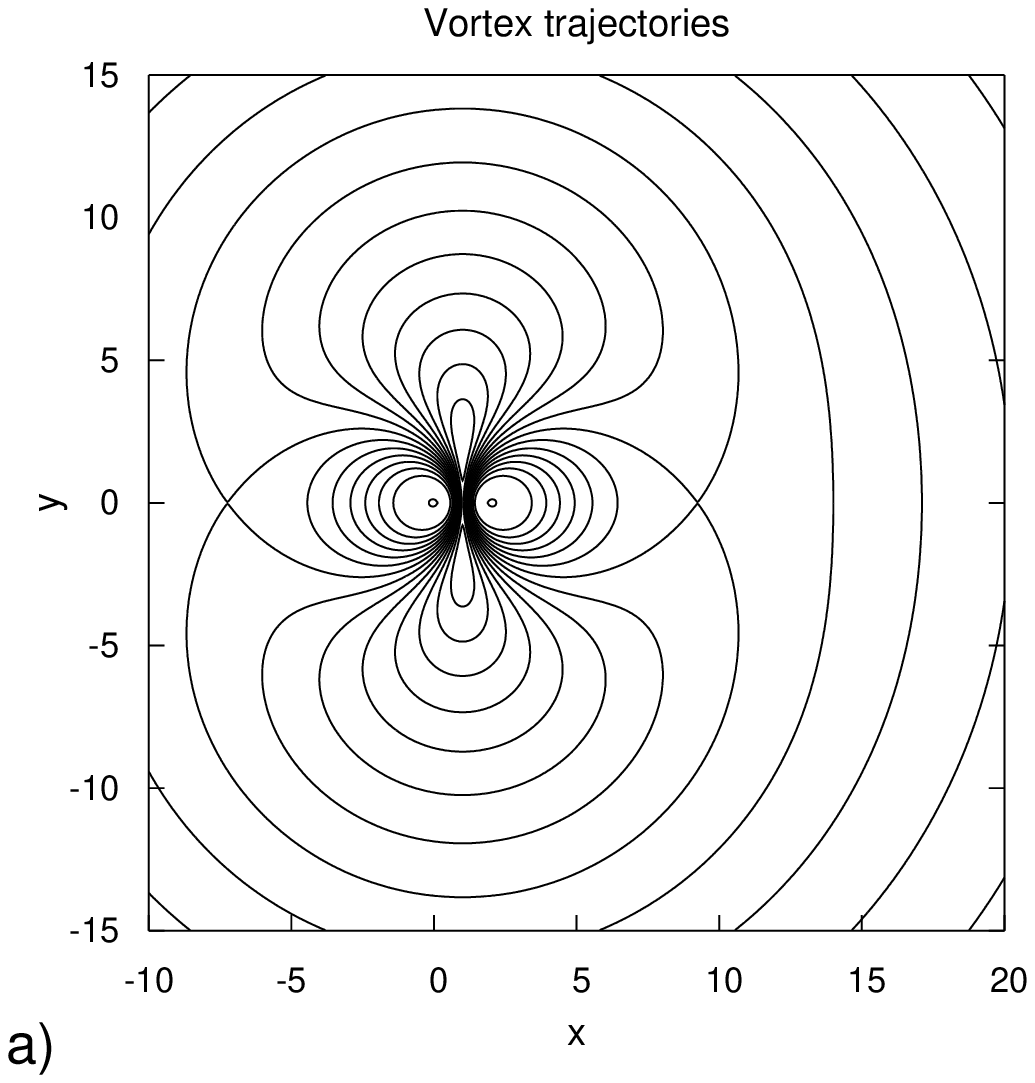, width=78mm}\\
 \epsfig{file=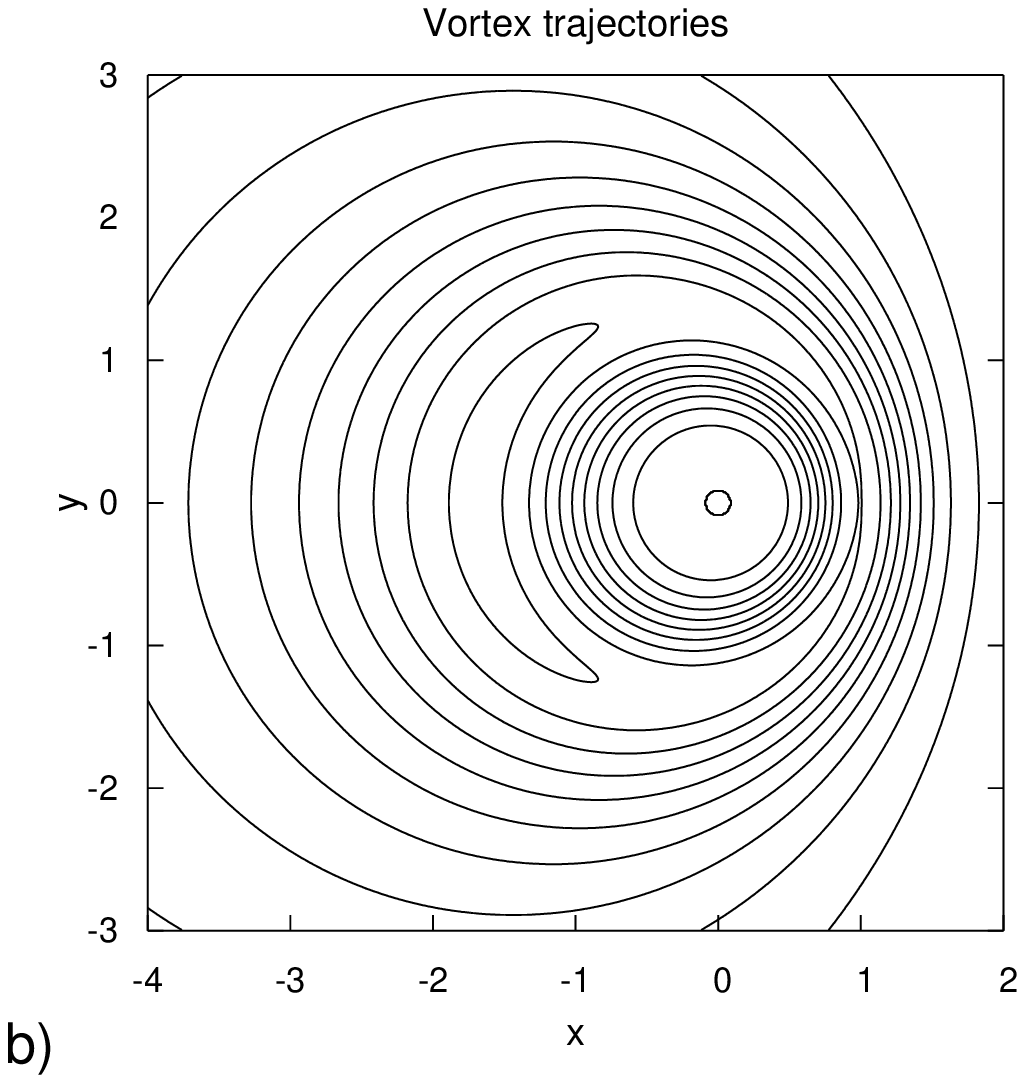, width=78mm}
\end{center}
\caption{Examples of the trajectories of a single vortex in the
plane for two sets of parameters: (a) $A=B=0.5$, $Q_0=1$, $Q_1=1$,
and (b) $A=0.8$, $B=0.2$, $Q_0=1$, $Q_1=0$.}
\label{portr_plane} 
\end{figure}

\begin{figure}
\begin{center}
a)\epsfig{file=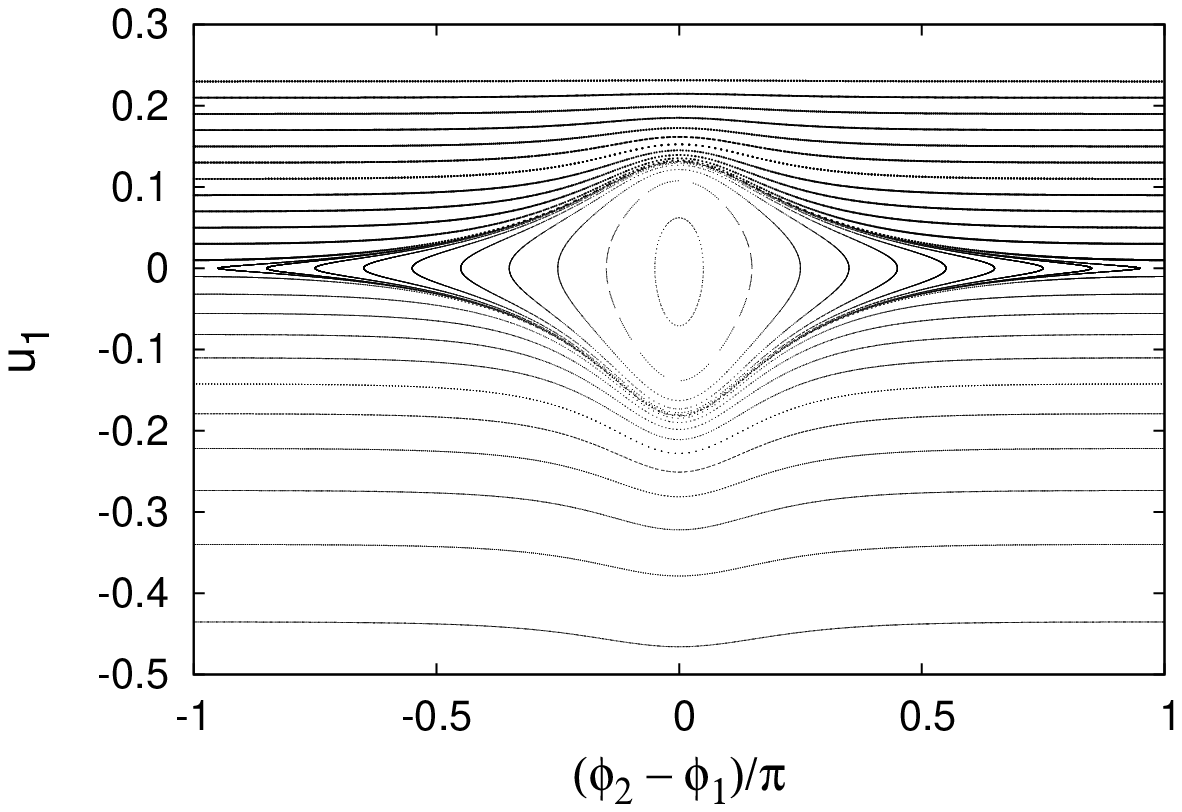, width=84mm}\\
b)\epsfig{file=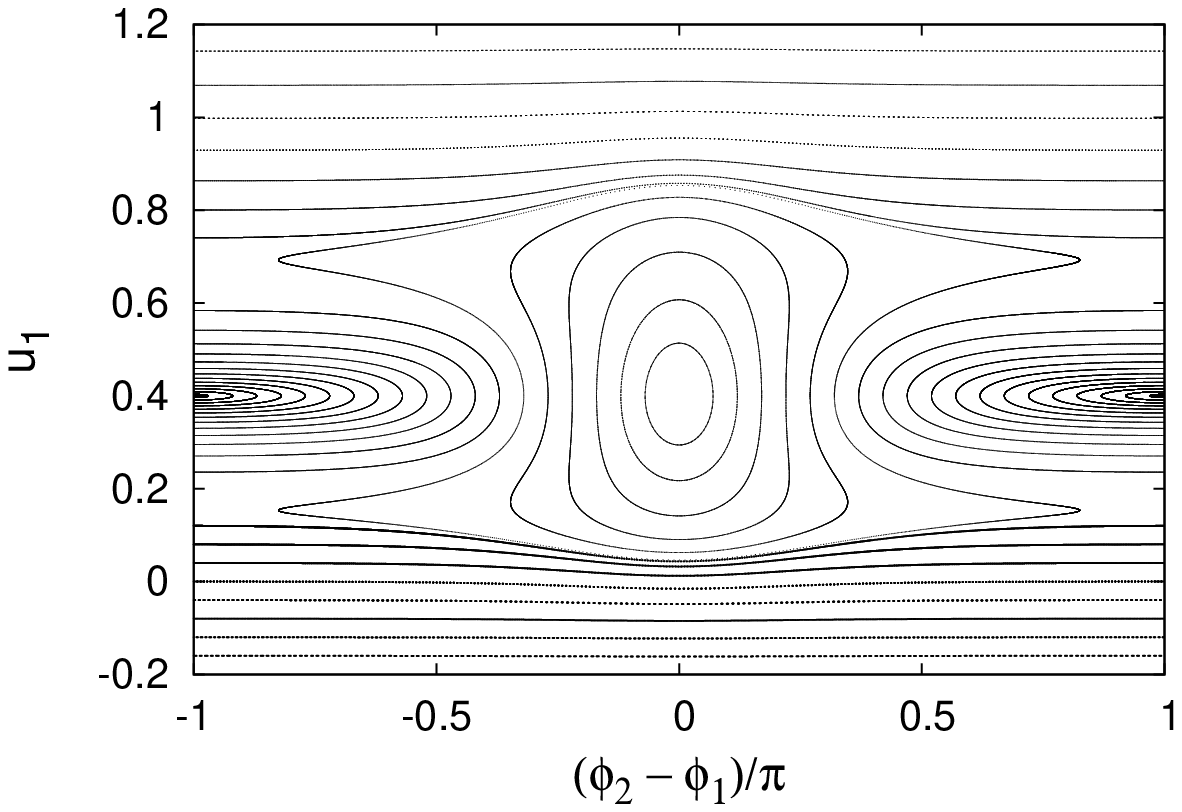, width=84mm}
\end{center}
\caption{Phase trajectories of two vortices at $Q_0 = 0$ and the
angular momentum (a) $M=2\mu(0)$, and (b) $M=2\mu(0.4)$.}
\label{N2Q0} 
\end{figure}

\begin{figure}
\begin{center}
\epsfig{file=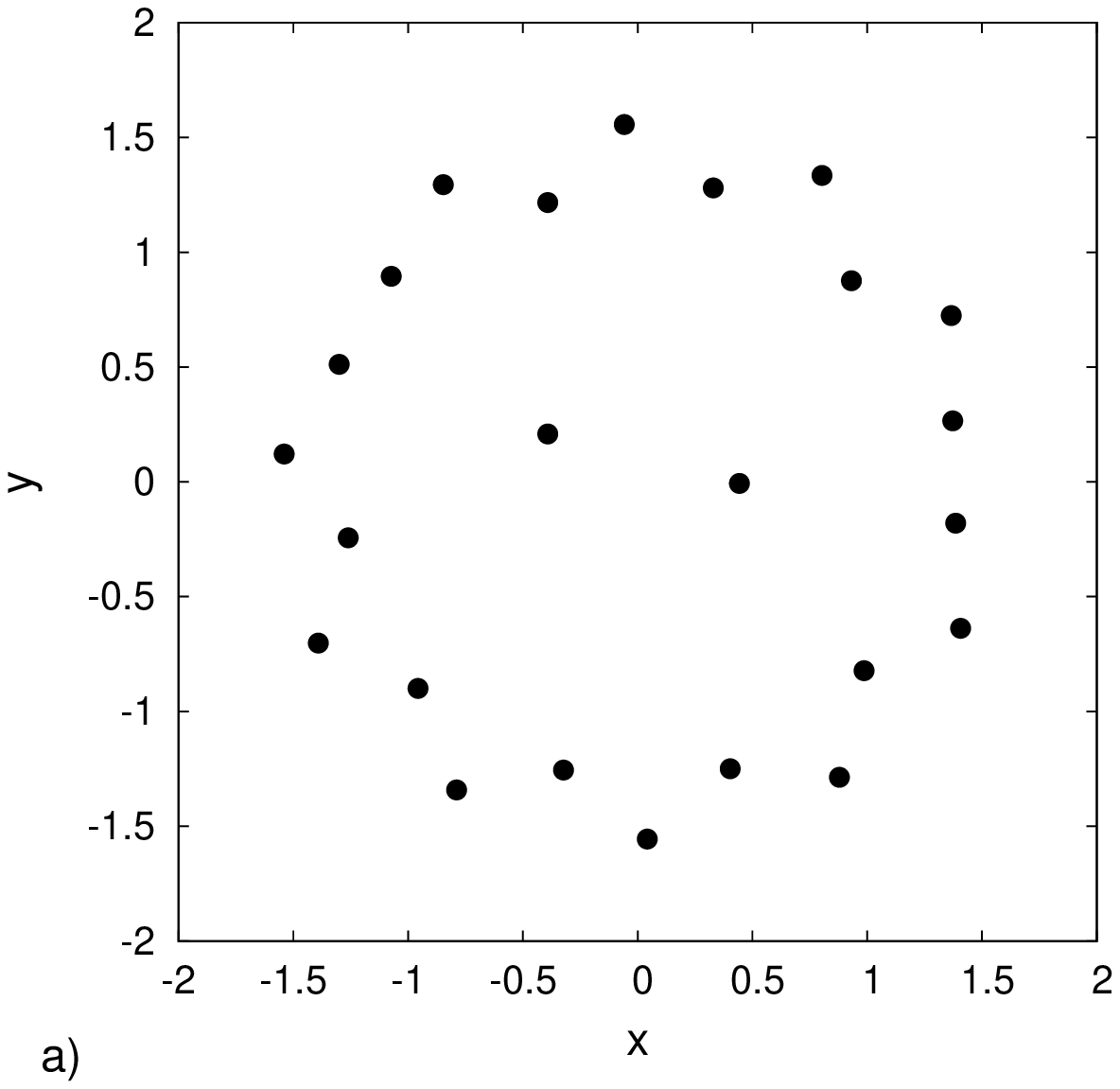, width=60mm}\\
\epsfig{file=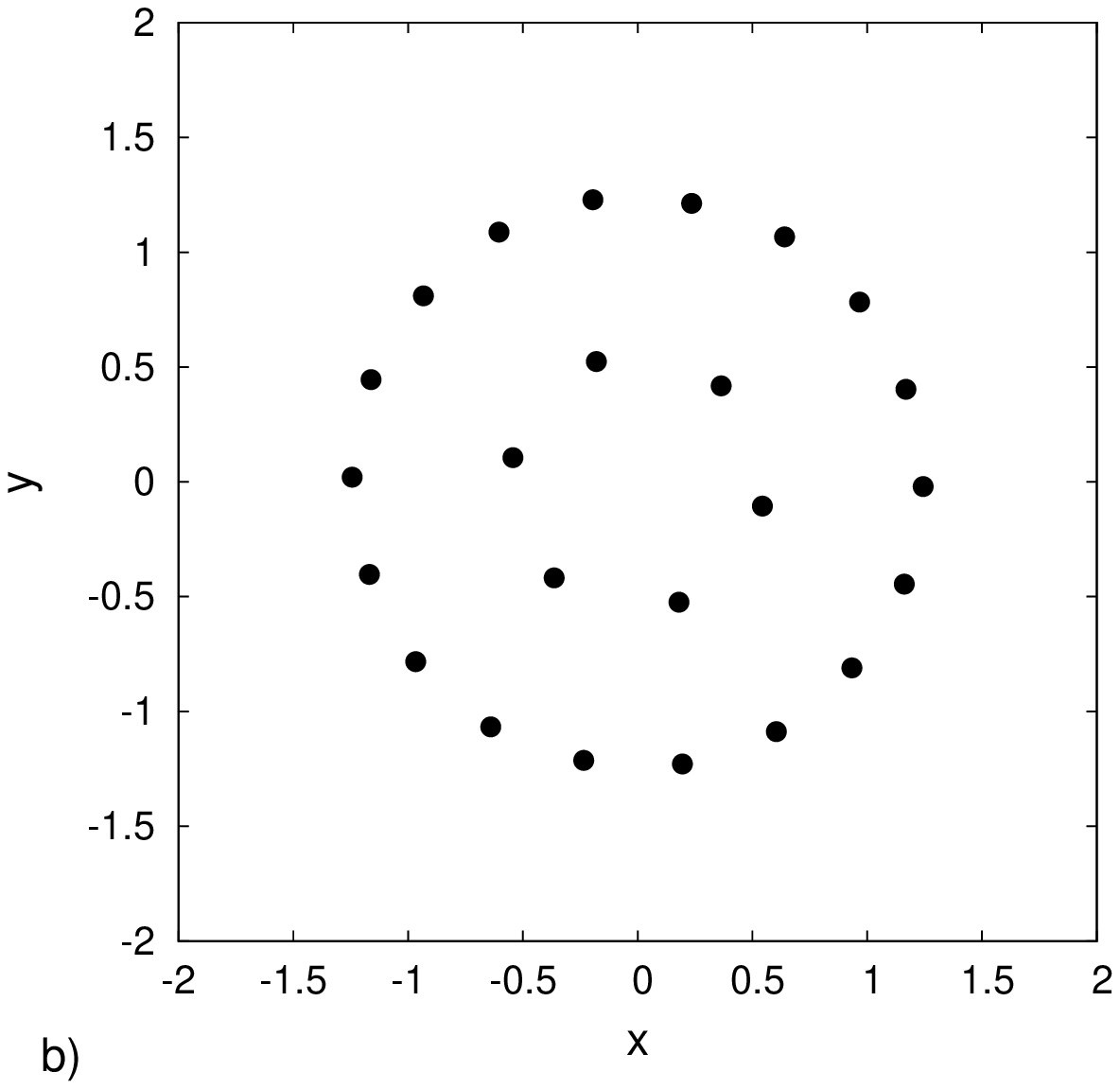, width=60mm}\\
\epsfig{file=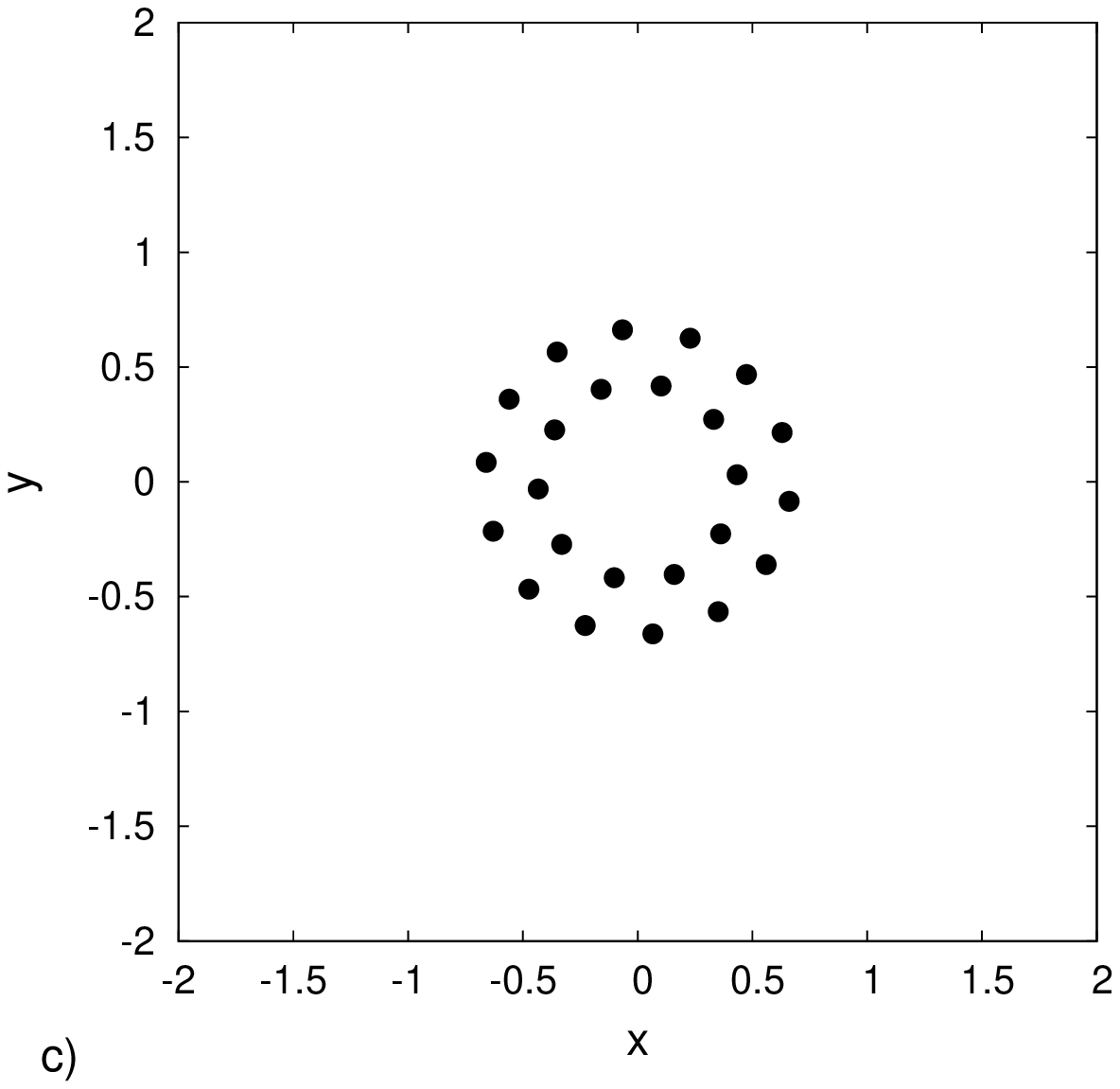, width=60mm}
\end{center}
\caption{Examples of steadily rotating configurations of $N=24$
vortices at $Q_0=12$ and the angular velocities: (a) $\Omega=15$, (b) $\Omega=20$, (c)  $\Omega=50$.}
\label{Omega} 
\end{figure}

\begin{figure}
\begin{center}
\epsfig{file=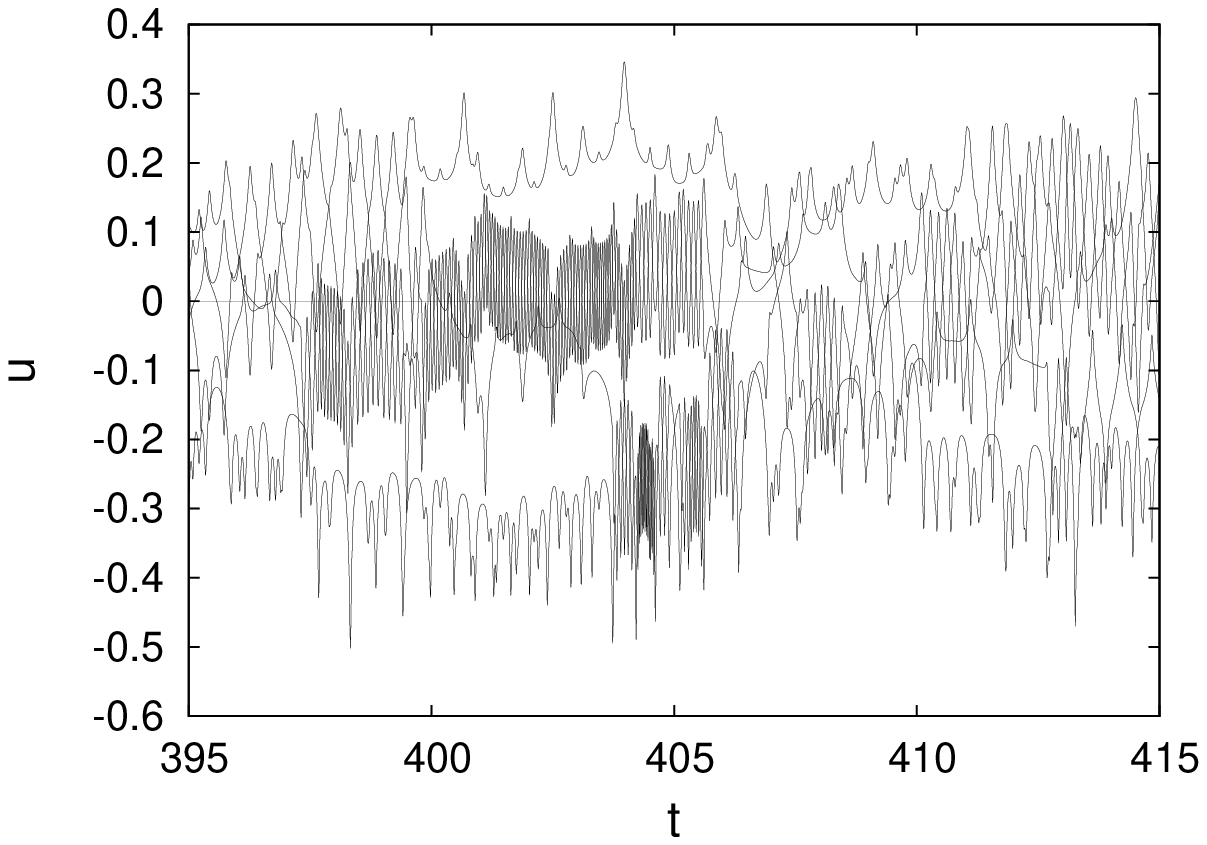, width=88mm}
\end{center}
\caption{Chaotic dynamics of five vortices at $Q_0=0$.}
\label{N5Q0-12} 
\end{figure}

\section{Axisymmetric case}

If $B=0$, the system becomes axisymmetric: the Jacobian $J(u)$ is independent of the angular coordinate $v=\phi$,
and the vortex Hamiltonian (22) is invariant with respect to a simultaneous shift of all variables
$v_n$. Therefore, one more integral of motion emerges, the angular momentum
\begin{equation}
M=\sum_n \sigma_n\mu(u_n),\qquad\mu(u)=\int\limits_{-\infty}^u J(s)\rho(s) d s. 
\end{equation}
Without loss of generality, we can set $A=1$ and then
\begin{equation}
\mu(u)=\int\limits_{-\infty}^{u}\frac{e^{2s}d s}{\cosh^2(\alpha s)}
=-\frac{2}{\alpha}\frac{e^{2u}}{1+e^{2\alpha u}}+\frac{2}{\alpha}\int\limits_0^{e^{2u}}\frac{d\chi}{1+\chi^\alpha}.
\end{equation}
At rational values of the parameter $\alpha$, the integral can be expressed in terms of elementary functions.

The presence of the second conservation law makes the problem of motion of two vortices integrable. The
projections of the phase trajectories of the system of two vortices on the $(\phi_2-\phi_1, u_1)$ plane at a constant 
$M$ value are exemplified in Fig.2, which implies, in particular, that the vortices located diametrically with
respect to the origin near the density maximum at $Q_0=0$ are at an unstable relative equilibrium.

In addition, ``rigid'' rotating configurations of $N$  vortices corresponding to local minima of the function 
$\tilde H_\alpha^{\{N\}}+\Omega M$, where $\Omega$ is the angular velocity of rotation, become possible at the 
axial symmetry. The examples are shown in Fig.3. As is seen, the vortices are not always situated near concentric 
circles at equilibrium. It should also be mentioned that, with an increase in the rotation rate, the vortices leave 
the maximum-density region and accumulate inside the ring, which agrees qualitatively with the formation of a
giant vortex.

The dynamics at $N\ge 3$ in a wide range of initial values exhibits the features of chaos. Chaos is caused
by the formation of pretty tight and thus rapidly rotating vortex pairs (of the same sign). The very fact of the
formation of a pair, its lifetime, size, rotation phase, and other characteristics appear to be almost unpredictable 
owing to the interaction with other vortices. The numerical example of the behavior of five vortices
is presented in Fig.4.

\section{Vortices on a curved surface}

One can imagine the situation where the three-dimensional trap potential forms a narrow curved
``canyon'', so that the Bose-Einstein condensate at equilibrium is not (inhomogeneously) planar, as was
assumed so far, but forms a relatively narrow layer near some curved surface in the three-dimensional space
[15-17]. In particular, ellipsoidal shells were prepared in the experiment reported in [18]. An interesting 
generalization of the above theory is the description of the dynamics of quantum vortices in such systems if the
variables $u$ and $v$  are regarded as the conformal coordinates on the curved surface ${\bf r}={\bf S}(u,v)$. 
It is easily verified that the equations of motion (13) and (14) for $u_n$ and $v_n$
preserve its structure and so does Hamiltonian (22) (with a possible restriction of $Q_\infty=0$,
depending on the type of the surface). An essential difference is that the function $J({\bf u})$ is now not 
the squared absolute value of an analytic function $z'(w)$ but a quite arbitrary conformal factor. 
In other words, the metric of the surface is
\begin{equation}
(d{\bf S})^2=J({\bf u})(du^2+dv^2).
\end{equation}

We consider as an example a unit sphere whose stereographic projection on the plane is determined by the formula
\begin{equation}
z + c=\tan(\Theta/2)e^{i\Phi},
\end{equation}
where $\Theta$ is the polar angle on the sphere, $\Phi$ is the azimuthal angle, and $c$  is an arbitrary complex 
constant. The conformal factor in terms of the variables ${\bf x}$ is
\begin{equation}
{\mathsf J}({\bf x})={4}/{(1+|x+iy +c|^2)^2}.
\end{equation}
It is worth mentioning that exactly this factor must be added to the left-hand sides of Eqs.(1) if one chooses
the variables $x$ and $y$ for the description of the vortex dynamics on a sphere. Accordingly, the logarithmically 
diverging terms in the Hamiltonian must be regularized with the use of ${\mathsf J}({\bf x})$.

The conformal factor in terms of ${\bf u}$  is the product ${\mathsf J}({\bf x}({\bf u}))\cdot|z'(w)|^2$,
which yields in our case
\begin{equation}
J({\bf u})=\frac{4A^2e^{-2u}}{\{|Ae^{-w}+B|^2+|c(Ae^{-w}+B)+1|^2\}^2}.
\label{J_sphere}
\end{equation}
Since this expression is finite everywhere, all vortices on the sphere at finite $u$ values are mobile, and hence
$Q_\infty =0$ should be in Hamiltonian (22).

Formula (28) is greatly simplified at $B=0$ and $c=0$, i.e., in the presence of the axial symmetry:
$J({\bf u})={1}/{\cosh^2(u-u_{\rm eq})}$, with $u=u_{\rm eq}$ corresponding
to the equator. As in the planar case, the axial symmetry implies the conservation of the angular momentum.

It is interesting that, even in the case of $u_{\rm eq}=0$, where the two-dimensional density profile is symmetric 
with respect the equatorial plane, the Hamiltonian of an odd number of vortices still is not symmetric with 
respect to the inversion $u_n\rightarrow -u_n$, since the velocity circulation quanta around the poles cannot be
equal to each other owing to the condition $Q_{\rm North}+ Q_{\rm South}+\sum_n\sigma_n=0$.

\section{Conclusions}

To summarize, a new, relatively simple, mathematically convenient, and rather rich model has been proposed 
which allows advancing quite far in the understanding of the mechanics of vortices in spatially inhomogeneous 
two-dimensional systems. A number of pictorial numerical examples have been presented.
Many other situations approximately corresponding to various actual experiments can be studied by varying
the parameters of the model. In particular, nonlinear oscillations of many vortices near steady configurations 
and the dynamics of oppositely oriented vortices remained beyond the scope of this publication.

Seemingly, many qualitative properties of the model Green's function found in this work persist for a wider class of 
the dependences $\rho(u)$, when $\alpha^2\neq\mbox{const}$ in Eq.(18). This question requires a separate investigation.


\begin{thebibliography}{99}



\bibitem{SF2000} A. A. Svidzinsky and A. L. Fetter,
Phys. Rev. A {\bf 62}, 063617 (2000).

\bibitem{FS2001} A. L. Fetter and A. A. Svidzinsky,
J. Phys.: Condens. Matter {\bf 13}, R135 (2001).

\bibitem{A2002} J. R. Anglin, Phys. Rev. A {\bf 65}, 063611 (2002).

\bibitem{SR2004} D. E. Sheehy and L. Radzihovsky, Phys. Rev. A {\bf 70}, 063620 (2004).

\bibitem{F2009}  A. L. Fetter, Rev. Mod. Phys. {\bf 81}, 647 (2009).

\bibitem{RP1994} B. Y. Rubinstein and L. M. Pismen, Physica D {\bf 78}, 1 (1994).

\bibitem{R2001} V. P. Ruban, Phys. Rev. E {\bf 64}, 036305 (2001).

\bibitem{R2017} V. P. Ruban, JETP {\bf 124}(6), ? (2017); arXiv:1612.00165.

\bibitem{gv1} K. Kasamatsu, M. Tsubota, and M. Ueda, Phys. Rev. A {\bf 66}, 053606 (2002).

\bibitem{gv2} S. Gupta, K. W. Murch, K. L. Moore, T. P. Purdy, and D. M. Stamper-Kurn,
Phys. Rev. Lett. {\bf 95}, 143201 (2005).

\bibitem{gv3} H. Fu and E. Zaremba, Phys. Rev. A {\bf 73}, 013614 (2006).

\bibitem{gv4} C. Ryu, M. F. Andersen, P. Clad\'e, V. Natarajan, K. Helmerson, and W. D. Phillips,
Phys. Rev. Lett. {\bf 99}, 260401 (2007).

\bibitem{gv5} A. Ramanathan, K. C. Wright, S. R. Muniz, M. Zelan, W. T. Hill, III, C. J. Lobb, 
K. Helmerson, W. D. Phillips, and G. K. Campbell, Phys. Rev. Lett. {\bf 106}, 130401 (2011).

\bibitem{gv6} S. Moulder, S. Beattie, R. P. Smith, N. Tammuz, and Z. Hadzibabic,
Phys. Rev. A {\bf 86}, 013629 (2012).

\bibitem{shell-1} O. Zobay and B. M. Garraway, Phys. Rev. Lett. {\bf 86}, 1195 (2001).

\bibitem{shell-2} O. Zobay and B. M. Garraway, Phys. Rev. A {\bf 69}, 023605 (2004).

\bibitem{shell-3} T. Fernholz, R. Gerritsma, P. Kr\"uger, and R. J. C. Spreeuw,
Phys. Rev. A {\bf 75}, 063406 (2007).

\bibitem{shell-4} B. E. Sherlock, M. Gildemeister, E. Owen, E. Nugent, and C. J. Foot,
Phys. Rev. A {\bf 83}, 043408 (2011). 

\bibitem{vs1} D. G. Dritschel and S. Boatto, Proc. R. Soc. A {\bf 471}: 20140890 (2015).

\bibitem{vs2} P. K. Newton, Theor. Comput. Fluid Dyn. {\bf 24}, 137 (2010).

\bibitem{vs3} R. B. Nelson and N. R. McDonald, Theor. Comput. Fluid Dyn. {\bf 24}, 157 (2010).

\bibitem{vs4} Y. Kimura, Proc. R. Soc. A {\bf 455}, 245 (1999).

\bibitem{vs5} A. Surana and D. Crowdy, J. Comput. Phys. {\bf 277}, 6058 (2008).

\bibitem{vs6} A. M. Turner, V. Vitelli, and D. R. Nelson, Rev. Mod. Phys. {\bf 82}, 1301 (2010).

\end{thebibliography}
\end{document}